\documentclass[showpacs,twocolumn]{revtex4}

\usepackage{amsmath,amssymb,mathrsfs}
\usepackage{latexsym}
\usepackage{graphicx}
\usepackage{epstopdf}

\newcommand{\bs}[1]{\boldsymbol{#1}}

\newcommand{\vac}{\left|\,0\,\right\rangle}
\newcommand{\ket}[1]{\left|#1\right\rangle}

\newcommand{\nn}{\nonumber}

\def\ie{\emph{i.e.},\ }
\def\eg{\emph{e.g.}\ }
\def\ea{\emph{et al.}}

\allowdisplaybreaks[1]

\begin{document}

\title{Exact models for trimerization and tetramerization in spin chains} 
\author{Stephan Rachel and Martin Greiter}
\affiliation{Institut f\"ur Theorie der Kondensierten Materie,
  Universit\"at Karlsruhe\\
  Postfach 6980, 76128 Karlsruhe, Germany} 
\pagestyle{plain} 
 \begin{abstract}
   We present exact models for an antiferromagnetic $S=1$ spin chain
   describing trimerization as well as for an antiferromagnetic
   $S=3/2$ spin chain describing tetramerization. These models can be
   seen as generalizations of the Majumdar--Ghosh model. For both
   models, we provide a local Hamiltonian and its exact three- or
   four--fold degenerate ground state wavefunctions, respectively.  We
   numerically confirm the validity of both models using exact
   diagonalization and discuss the low lying excitations.
 \end{abstract}


\maketitle

\section{Introduction}
Spin 1 antiferromagnetic chains have been the subject of extensive
study in the 1980s and 90s. Many studies were motivated by Haldane's
identification of the O(3) non--linear sigma model as the effective
low--energy theory for spin chains~\cite{haldane83pla464}.  He pointed
out that antiferromagnetic spin chains with integer spin
representations possess a finite energy gap in the excitation spectrum
and that the ground state correlations exhibit an exponential decay.
Haldane's conjecture was substantiated through a rigorous theorem by
Affleck and Lieb~\cite{affleck-86lmp57} and through the AKLT
model~\cite{affleck-87prl799,arovas-88prl531,kluemper-93epl293} on
theoretical site, and through the observation of the Haldane gap in
spin 1 chains on experimental
site~\cite{buyers-86prl371,renard-87epl945,katsumata-89prl86,ajiro-89prl1424,ma-92prl3571}. Haldane's
prediction has also been confirmed in detail by numerical studies~\cite
{nightingale-86prb659,takahashi89prl2313,white-93prb3844}.

Recently, spin 1 chains have seen a considerable renewal of interest.
Experimental realizations using polar molecules stored in optical
lattices have been proposed~\cite{micheli-06np341}, with the spin
represented by a single-valence electron of a heteronuclear molecule.
Moreover, it appears possible that the one--parameter family of
Hamiltonians
\begin{equation}
\label{ham.spin1.general}
\mathcal{H}_{\theta}=\sum_{i=1}^N \cos{\theta}\,\bs{S}_i\bs{S}_{i+1} + 
\sin{\theta} \left( \bs{S}_i\bs{S}_{i+1} \right)^2
\end{equation}
can be engineered in optical lattices using
cold spin 1 bosonic particles with antiferromagnetic interactions,
such as $^{23}$Na, for arbitrary values of
$\theta$~\cite{yip03prl250402,imambekov-pra03063602}. 

The phase diagram of the model \eqref{ham.spin1.general} as a
function of $\theta$ has been investigated by numerous authors (\eg
\cite{
affleck86npb409,papanicolaou88npb367,parkinson88jpc3793,
barber-89prb4621,sorensen-90prb754,fath-91prb11836,
chubukov91prb3337,fath-93prb872,xiang-93prb303,fath-95prb3620,
schollwock-96prb3304,pati-96epl707,kolezhuk-96prl5142,
itoi-97prb8295,laeuchli-06prb144426,karimipour-08prb094416}
and references therein) and is by now well understood.

\begin{figure}[t!]
\begin{center}
\includegraphics[scale=1.]{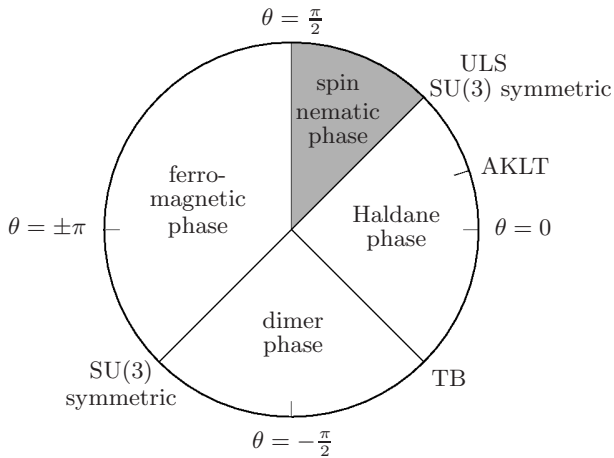}
\caption{Phase diagram of Hamiltonian \eqref{ham.spin1.general}
         as a function of $\theta$.}
 \label{fig:theta-circle}
\end{center}
\end{figure}

The point $\theta=0$ on the circle shown in
Fig.~\ref{fig:theta-circle}, the antiferromagnetic Heisenberg point,
is embedded in the so--called Haldane phase ($-\pi/4<\theta<\pi/4$)
which is characterized by a unique ground state, exponentially
decaying correlations, and a gap between the ground state and the
excited states.  The Haldane phase includes at
$\theta_{\textrm{VBS}}=\arctan{(1/3)}$ the valence bond solid (VBS) or
AKLT model. The AKLT Hamiltonian shares the most properties of the
isotropic Heisenberg Hamiltonian but, in contrast to the isotropic
Heisenberg model, possesses a ground state which can be written out
explicitly.

Above the Haldane phase in Fig. \eqref{fig:theta-circle}, there is a
critical phase ($\pi/4<\theta<\pi/2$) with spin nematic
correlations~\cite{laeuchli-06prb144426}.  The phase transition at
$\theta_{\textrm{ULS}}=\pi/4$ was proposed to be of
Kosterlitz--Thouless type~\cite{fath-93prb872,itoi-97prb8295}.  At the
transition point, the Hamiltonian \eqref{ham.spin1.general} reduces to
the Uimin--Lai--Sutherland (ULS)
model~\cite{uimin70jetp225,lai70jmp1675,sutherland75prb3795} which
 exhibits explicit SU(3) symmetry.  The ULS
model is a sum of permutation operators and exactly solvable via the
nested Bethe
ansatz. 

At $\theta=\pi/2$, the Hamiltonian \eqref{ham.spin1.general} becomes
ferromagnetic with gapless excitations. It reaches the ferromagnetic
Heisenberg point at $\theta=\pm\pi$ and undergoes a first order phase
transition to a dimerized phase at $\theta=-3\pi/4$ where
\eqref{ham.spin1.general} is again SU(3) symmetric and has a highly
degenerate ground state~\cite{batista-02prb180402}. Close to this
point there was a long--standing discussion regarding the possible
existence of a small spin nematic phase. Recently, this was ruled out
by numerical and analytical
arguments~\cite{laeuchli-06prb144426,grover-07prl247202}. In the
dimerized phase ($-3\pi/4 < \theta<-\pi/4$), the excitations are
gaped. At the Takhtajan--Babudjan point $\theta_{\rm{TB}}=-\pi/4$, the
gap closes and the model is again exactly solvable via the nested
Bethe ansatz~\cite{takhtajan82pla479,babudjian}, has gapless
excitations, and a unique ground state. Finally, the phase transition
to the Haldane phase at $\theta=-\pi/4$ is of second
order~\cite{affleck85prl1355,affleck86npb409}.

Fath and S\'olyom~\cite{fath-91prb11836,fath-93prb872} observed in 1991
a {\it period tripling} in the spectrum of \eqref{ham.spin1.general}
in the critical phase ($\pi/4 < \theta < \pi/2$). The observation of
{\it three soft modes} in their numerical studies caused a controversy
whether or not it is a trimer
phase~\cite{xian93jpcm7489,reed94jpamg69,bursill-95jpamg2109}.
Subsequent numerical studies concluded that there is no trimer
phase~\cite{schmitt-98prb5498}. Recently, it was found that the
dominant correlations in this phase are not of singlet, but of spin
nematic (quadrupolar) character~\cite{laeuchli-06prb144426}.

Regardless of the spin nematic phase, $S=1$ models yielding
trimerization can be constructed with a spin interaction beyond the
nearest--neighbor case. S\'olyom and Zittartz~\cite{solyom-00epl389}
presented such a model with four--site interaction. In this model, the
trimer singlets are nested and each trimer-singlet is placed on three
non--neighboring sites, say on sites $3i-2$, $3i$, and $3i+2$
($i=1,\ldots,N/3$). Most recently, Corboz
\ea~\cite{corboz-07prb220404} investigated numerically the
bilinear--biquadratic Heisenberg model with nearest- and next nearest
neighbor interactions:
\begin{equation}
  \label{bilinear-biquadratic-HM}
\begin{split}
\mathcal{H}=\,&J_1\sum_{i=1}^N \cos{\theta}\,\bs{S}_i\bs{S}_{i+1} + 
\sin{\theta} \left( \bs{S}_i\bs{S}_{i+1} \right)^2\\[0pt]
+&J_2\sum_{i=1}^N \cos{\theta}\,\bs{S}_i\bs{S}_{i+2} + 
\sin{\theta} \left( \bs{S}_i\bs{S}_{i+2} \right)^2.
\end{split}\end{equation}
For certain values of the ratio $J_2/J_1$, they found in a small
region around $\theta_{{\rm{ULS}}}$ a trimerized phase in which the
ground state becomes three--fold degenerate as they approach the
thermodynamic limit. This can be seen as the analogy of the dimer
phase in the $S=1/2$ $J_1$--$J_2$--model for $J_2/J_1 \gtrsim
0.2411$~\cite{julien-83baps34,okamoto-92pla433,eggert96prb9612}.
Recently, trimerization and tetramerization was also discussed by
Lecheminant and
Totsuka~\cite{lecheminant-06prb224426,lecheminant-06jsm12001} within a
field theoretical approach where they considered a self-dual SU($n$)
Sine--Gordon model.

In the first part of this article, we discuss an exact model for
trimerization in an $S=1$ antiferromagnet. The model is exact for
finite (and hence trivially also for infinite) chains. The ground
state consists of simple trimer singlets (each of which is placed on
three consecutive sites) and is three--fold degenerate. Since the
three ground states can be written as valence bond solids, the model
is in the spirit of the Majumdar--Ghosh (MG)
model~\cite{majumdar-69jmp1399,asoudeh-07prb064433} for $S=1/2$, where
dimers occupy pairs of neighboring sites. We present a local
Hamiltonian for which the trimer states are the exact zero--energy
ground states, and identify numerically the elementary triplet
excitation. It is a soliton consisting of two antisymmetrically
coupled spins on adjacent sites. In the second part, we discuss the
possible generalization to higher spin and introduce an exact model
for the tetramerized (or, alternatively, quadrumerized) phase in a
spin $S=3/2$ antiferromagnet. We numerically confirm the validity of
the model using exact diagonalization. We further discuss low--lying
excitations and give an outlook for tetramerization in other models.

\section{Model for trimerization}\label{sec:spin1model}

We consider a chain with $N=3\mu$ sites ($\mu$ integer) and impose
periodic boundary conditions (PBCs). To write the trimer ground states
as products of trimer singlets, we define an operator $T[i,j,k]$ which
creates a trimer singlet on sites $i$, $j$, and $k$ as follows:
\begin{equation}\label{trimer-singlet}
\begin{split}
  T[i,j,k]\,\vac=\frac{1}{\sqrt{6}}\Big( &\ket{+,0,-}+\ket{0,-,+}
  +\ket{-,+,0} \Big.\\
  \Big. - &\ket{+,-,0} - \ket{-,0,+} - \ket{0,+,-}\Big),
\end{split}
\end{equation}
where $\ket{+,0,-} =
c_{i,+}^{\dagger}c_{j,0}^{\dagger}c_{k,-}^{\dagger}\vac$ etc., with the
usual fermionic creation operators $c_{i,\alpha}^{\dagger},
\,\alpha = +,0,-$.  Using equation \eqref{trimer-singlet} the
three trimer ground states are given by
\begin{eqnarray}
\nn \ket{\psi_1} &=& \prod_{i=0}^{\mu-1} T[3i+1,3i+2,3i+3] \vac,\\
\label{trimer-groundstates}
\ket{\psi_2} &=& \prod_{i=0}^{\mu-1} T[3i+2,3i+3,3i+4] \vac,\\
\nn \ket{\psi_3} &=& \prod_{i=0}^{\mu-1} T[3i+3,3i+4,3i+5] \vac.
\end{eqnarray}

\begin{figure}[b!]
  \centering
  \vspace{3pt}
  \includegraphics[scale=0.8]{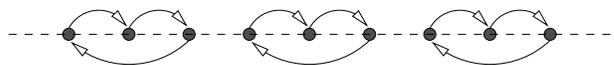}
  \caption{Illustration of one of the trimer ground states on a chain
    with $N=3\mu$ sites, where three neighboring spins are
    antisymmetrically coupled.}
  \label{fig:trimer-groundstate}
\end{figure}

Note that the trimer ground states \eqref{trimer-groundstates} break
translational symmetry spontaneously, while they are invariant under
translations by three lattice spacings. To illustrate the states, we
show one of them in Fig.~\ref{fig:trimer-groundstate} by connecting
the sites belonging to the trimer singlets by arrows.

To find an exact parent Hamiltonian for the trimer states,
interactions including first-, second-, and third-nearest neighbors
are required. More generally speaking, we conjecture that for the
construction of an exact Hamiltonian for ``$n$-merization'' consisting
of local projection operators, interactions involving $n$+1
neighboring sites are required. Assuming interactions involving less
than $n+1$ sites, additional non-$n$-merized states might occur which
will be annihilated by the Hamiltonian. We further stress that
SU($3n$) symmetry is a crucial requirement for trimerization (even
though we are unable to prove this statement rigorously).

For convenience we introduce the auxiliary operator
$\mathcal{X}^{(4)}$. It acts on four consecutive sites,
\begin{equation}\label{ham.auxiliary}
{\mathcal{X}}_i^{(4)}= \sum_{\underset{j<j'}{j,j' = i}}^{i+3} 
\bs{S}_j\bs{S}_{j'} + \left( \bs{S}_j\bs{S}_{j'} \right)^2, 
\end{equation}                                             
where $\bs{S}_j$ is the usual spin 1 operator.             
In terms of these the Hamiltonian we propose is given by
\begin{equation}
  \label{ham.trimer}
  \mathcal{H}=\sum_{i=1}^N \big( \,{\mathcal{X}}_i^{(4)} - 6 \,\big)
       \big( \,{\mathcal{X}}_i^{(4)} -4 \,\big).
\end{equation}
As in the case of the MG Hamiltonian, equation \eqref{ham.trimer} is a
sum over projection operators. The trimer ground states
\eqref{trimer-groundstates} will be annihilated by these operators,
\begin{equation}\nn
\big( \,{\mathcal{X}}_i^{(4)} - 6 \,\big)
\big( \,{\mathcal{X}}_i^{(4)} - 4 \,\big)\ket{\psi_{\nu}}=0
\end{equation}
for $\nu=1,2,$ or $3,$ and thus $\mathcal{H}\ket{\psi_{\nu}}=0$.

The key to the trimer phase is the explicit SU(3) symmetry at the ULS
point in Fig.\,\ref{fig:theta-circle}. The symmetry emerges because
the bilinear--biquadratic $S=1$ Heisenberg interaction becomes
proportional to the SU(3) symmetric permutation operator if the
bilinear and the biquadratic term appear with the same coefficients
(as it happens at the ULS point).  The SU(3) symmetric permutation
operator $\mathcal{P}^{(3)}_{\alpha\beta}$ fulfils
\begin{equation}\nn
  \begin{split}
 &\mathcal{P}^{(3)}_{12}\ket{+,0}=\ket{0,+},\quad
  \mathcal{P}^{(3)}_{12}\ket{+,-}=\ket{-,+},\\
 &\mathcal{P}^{(3)}_{12}\ket{0,-}=\ket{-,0},\quad
  \mathcal{P}^{(3)}_{12}\ket{+,+}=\ket{+,+}~\textrm{etc.},
  \end{split}
\end{equation}
\ie $\mathcal{P}^{(3)}_{\alpha\beta}$ permutes the spins on site
$\alpha$ and $\beta$. The permutation operator and the spin 1
operators are related by
\begin{equation}
\label{SU3-SU2-mapping}
\mathcal{P}^{(3)}_{\alpha\beta}~=~\bs{S}_{\alpha}\bs{S}_{\beta} + 
\left(\bs{S}_{\alpha}\bs{S}_{\beta}\right)^2 - 1.
\end{equation}
On the other hand, the
permutation operator is related to the SU(3) spin operators
$\bs{J}_{\alpha}$ by
\begin{equation}\label{su3-permuation-operator}
  \mathcal{P}^{(3)}_{\alpha\beta}~=~2\,\bs{J}_{\alpha}
\bs{J}_{\beta}+\frac{1}{3}.
\end{equation}
The SU(3) generators at each lattice site $\alpha$ are defined
as
\begin{equation}
  \nn \bs{J}_{\alpha}^{a}=\frac{1}{2}\sum_{\sigma,\sigma'=+,0,-}
  c_{\alpha\sigma}^{\dagger} \lambda_{\sigma \sigma'}^a 
c_{\alpha\sigma'}^{\phantom{\dagger}}, \quad a=1,\ldots,8,
\end{equation}
where $\lambda^a$ are the SU(3) Gell--Mann matrices (see
\eg~\cite{greiter-07prb184441}).  Equations \eqref{SU3-SU2-mapping}
and \eqref{su3-permuation-operator} allows us to define a spin $S=1$
model which is simultaneously an SU(3) model. The ground states of the
model are given by the trimer products \eqref{trimer-groundstates}.
Note that \eqref{ham.auxiliary} is up to an additive and
multiplicative constant equal to the Casimir of the total SU(3) spin
on four consecutive sites; the Hamiltonian \eqref{ham.trimer}
corresponds to equation (12) in Ref. \cite{greiter-07prb184441}.  We
now explain the explicit construction for the SU(3) Hamiltonian which
is equivalent to \eqref{ham.trimer} and which describes trimerization.
(A detailed discussion of this Hamiltonian for the SU(3) system can be
found in \cite{greiter-07prb184441}.)

We consider a spin chain with a fundamental representation $(1,0)$ of
SU(3) on each lattice site. As mentioned above, for exact
trimerization interactions involving four neighboring sites are
required. To find the relevant SU(3) representations appearing on four
consecutive sites, we couple four fundamental representations,
$(1,0)\otimes(1,0)\otimes(1,0)\otimes(1,0)=3\cdot (1,0)\oplus 2\cdot
(0,2)\oplus 3\cdot (2,1)\oplus(4,0)$. In the ground state, only the
representations $(1,0)$ and $(0,2)$ are present. This can be seen by
considering four neighboring sites in the ground state configuration
(see Fig.\,\ref{fig:trimer-groundstate}): either three spins are
antisymmetrically coupled to a singlet (\ie
$\mathcal{A}[\,(1,0)\otimes (1,0)\otimes (1,0)\,]\otimes (1,0)=(1,0)$)
or pairs of spins are antisymmetrically coupled to anti--fundamental
representations $(0,1)$ (\ie $\mathcal{A}[(1,0)\otimes (1,0)]\otimes
\mathcal{A}[(1,0)\otimes (1,0)]=(0,1)\otimes (0,1)=(1,0)\oplus
(0,2)$).  Hence, the projection operator onto the subspace $(2,1)$ and
$(4,0)$ applied to one of the trimer ground states must be zero,
\begin{equation}\label{project-su3}
  P\setlength{\unitlength}{5pt}
\begin{picture}(5,0)(0,0.5)
\linethickness{0.3pt}
\put(0,1){\line(1,0){4}}
\put(0,0){\line(1,0){4}}
\put(0,0){\line(0,1){1}}
\put(1,0){\line(0,1){1}}
\put(2,0){\line(0,1){1}}
\put(3,0){\line(0,1){1}}
\put(4,0){\line(0,1){1}}
\end{picture}(i,j,k,l)\,\ket{\psi_{\nu}}=
P\setlength{\unitlength}{5pt}
\begin{picture}(4,0)(0,0.5)
\linethickness{0.3pt}
\put(0,1){\line(1,0){3}}
\put(0,0){\line(1,0){3}}
\put(0,-1){\line(1,0){1}}
\put(0,-1){\line(0,1){2}}
\put(1,-1){\line(0,1){2}}
\put(2,0){\line(0,1){1}}
\put(3,0){\line(0,1){1}}
\end{picture}
(i,j,k,l)\,\ket{\psi_{\nu}}=0
\end{equation}
for $\nu=1,2,$ or $3$ if $i$, $j$, $k$, and $l$ label four consecutive
sites. The Young tableaux $\setlength{\unitlength}{5pt}
\begin{picture}(4.5,0)(0,-0.0)
\linethickness{0.3pt}
\put(0,1){\line(1,0){4}}
\put(0,0){\line(1,0){4}}
\put(0,0){\line(0,1){1}}
\put(1,0){\line(0,1){1}}
\put(2,0){\line(0,1){1}}
\put(3,0){\line(0,1){1}}
\put(4,0){\line(0,1){1}}
\end{picture}$ corresponds to the representation $(4,0)$ and
$\setlength{\unitlength}{5pt}
\begin{picture}(3.5,0)(0,-0.75)
\linethickness{0.3pt}
\put(0,1){\line(1,0){3}}
\put(0,0){\line(1,0){3}}
\put(0,-1){\line(1,0){1}}
\put(0,-1){\line(0,1){2}}
\put(1,-1){\line(0,1){2}}
\put(2,0){\line(0,1){1}}
\put(3,0){\line(0,1){1}}
\end{picture}$ to the representation $(2,1)$.
The conditions \eqref{project-su3} singles out the states 
\eqref{trimer-groundstates} uniquely as ground states, which enables 
us to write the parent Hamiltonian as
\begin{equation}
  \label{trimer.projection}
  \mathcal{H} 
=\sum_{\langle ijkl \rangle} 
\Bigg(P\setlength{\unitlength}{5pt}
\begin{picture}(5,0)(0,0.5)
\linethickness{0.3pt}
\put(0,1){\line(1,0){4}}
\put(0,0){\line(1,0){4}}
\put(0,0){\line(0,1){1}}
\put(1,0){\line(0,1){1}}
\put(2,0){\line(0,1){1}}
\put(3,0){\line(0,1){1}}
\put(4,0){\line(0,1){1}}
\end{picture}(i,j,k,l)+P\setlength{\unitlength}{5pt}
\begin{picture}(4,0)(0,0.5)
\linethickness{0.3pt}
\put(0,1){\line(1,0){3}}
\put(0,0){\line(1,0){3}}
\put(0,-1){\line(1,0){1}}
\put(0,-1){\line(0,1){2}}
\put(1,-1){\line(0,1){2}}
\put(2,0){\line(0,1){1}}
\put(3,0){\line(0,1){1}}
\end{picture}
(i,j,k,l)\Bigg).
\end{equation}
The brackets $\langle \cdot \rangle$ indicate summation over four
neighboring sites along the chain. We can rewrite this Hamiltonian in
terms of SU(3) spin operators, such that it annihilates the states
which carry exclusively the representation $(1,0)$ or $(0,2)$ on four
consecutive sites:
\begin{equation}\label{spinhamiltonian-su3}\begin{split}
  \mathcal{H}=\sum_{i=1}^N&\left(\big(\bs{J}_i+ \bs{J}_{i+1}
  +\bs{J}_{i+2}+\bs{J}_{i+3}\big)^2-\frac{4}{3}\right)\\
&\times\left(\big(\bs{J}_i+ \bs{J}_{i+1}
  +\bs{J}_{i+2}+\bs{J}_{i+3}\big)^2-\frac{10}{3}\right),
\end{split}
\end{equation}
where we have used that the eigenvalues of the squared total spin on
four neighboring sites, $\bs{J}_{{\rm 4\,sites}}^2$, is $\frac{4}{3}$
in case of the representation $(1,0)$ and $\frac{10}{3}$ in case of
the representation $(0,2)$. 
The eigenvalues of the Casimir operators for the representations
$(2,1)$ or $(4,0)$ are both larger than $\frac{10}{3}$. The
Hamiltonian \eqref{spinhamiltonian-su3} annihilates the ground states
\eqref{trimer-groundstates} while the other states end up with a
positive energy. Finally, we replace the SU(3) spin operators via
\eqref{su3-permuation-operator} and \eqref{SU3-SU2-mapping} by spin
$S=1$ operators to obtain \eqref{ham.trimer}.

\begin{figure}[t!]
  \centering
  \includegraphics[scale=0.685,angle=0]{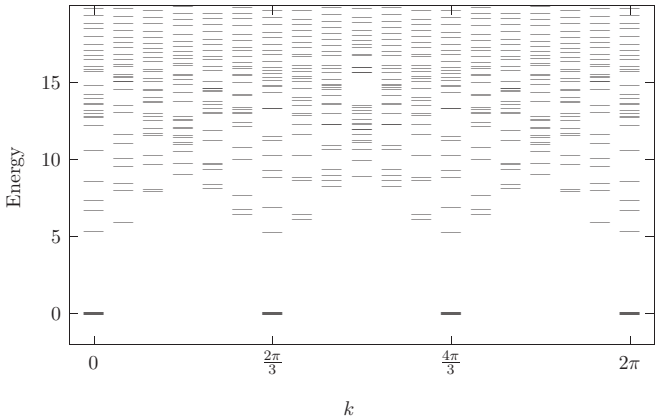}
  \caption{Spectrum of the spin 1 model \eqref{ham.trimer} for
    trimerization on a chain with $N=18$ sites. The zero energy ground
    states (labeled by thick lines in the spectrum) are at
    $k=0,\frac{2\pi}{3}, \frac{4\pi}{3}$ in the Brioullin zone.}
  \label{fig:spectrum}
\end{figure}

We have confirmed our predictions with exact diagonalization (ED) for
chains with $N=9$ and $N=12$ sites. In both cases we found precisely
three zero--energy ground states. These ground states are in the
Brioullin zone located at $k=0,\frac{2\pi}{3},\frac{4\pi}{3}$,
reflecting the fact that the system is translationally invariant under
translations by three lattice spacings. We have plotted the spectrum
$E(k)$ in Fig.~\ref{fig:spectrum} for $N=18$ sites where we used
Lanczos routine for diagonalization.

\begin{figure}[b]
  \centering
\includegraphics[scale=1.]{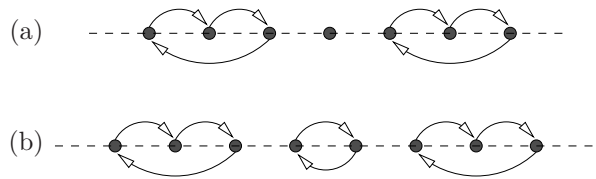}
  \caption{(a) Possible triplet excitation as a domain wall on a chain
    with $3\mu+1$ sites. 
    (b) Possible triplet excitation as a domain wall on a chain with
    $3\mu+2$ sites. The excitations consists of two antisymmetrically
    coupled spins on adjacent sites.}
  \label{fig:3-exc}
\end{figure}

Even though we cannot write down any of the excited states exactly, we
are able to elaborate on key properties like the quantum numbers
involved.  To create an excitation, we inevitably have to break a
trimer, \ie create a domain wall between the degenerate ground states.
There are, however, two different types of domain walls (see
Fig.\,\ref{fig:3-exc} (a) and (b)) which correspond to
different excitations.

As explained above, our model exhibits an SU(3) symmetry, and we may
view the trimer singlets \eqref{trimer-groundstates} as SU(3)
singlets. Breaking such an SU(3) singlet yields either an individual
SU(3) spin with fundamental representation $(1,0)$ or two
antisymmetrically coupled spins on adjacent sites with the resulting
anti-fundamental representation $(0,1)$, \ie
$\mathcal{A}[(1,0)\otimes(1,0)]=(0,1)$. Note that both $(1,0)$ and
$(0,1)$ are three--dimensional representations.  Since the only
three--dimensional representation of SU(2) is the triplet, the
excitation has to be a triplet regardless of which type it corresponds
to.

The first type consists of an individual spin $S=1$, playing the role
of a domain wall between two different trimer ground states, \eg
between $\ket{\psi_1}$ and $\ket{\psi_2}$ (see Fig.\,\ref{fig:3-exc}\,(a)).

The second type consists of two spins $S=1$ on adjacent sites. They
are coupled such that the internal Hilbert space spanned by this
excitation is three dimensional, \ie
$\ket{e_+}=\frac{1}{\sqrt{2}}\left(\ket{+,0}-\ket{0,+}\right)$,
$\ket{e_0}=\frac{1}{\sqrt{2}}\left(\ket{+,-}-\ket{-,+}\right)$, and
$\ket{e_-}=\frac{1}{\sqrt{2}}\left(\ket{0,-}-\ket{-,0}\right)$.  This
second type of triplet excitation is also as a domain wall between two
different ground states, \eg between $\ket{\psi_1}$ and $\ket{\psi_3}$
(see Fig.\,\ref{fig:3-exc}\,(b)). Since either type of domain wall
could in principle decay into two domain walls of the other type, only
one type can be an 
eigenstate of the Hamiltonian \eqref{ham.trimer} above.


In Ref.~\cite{greiter-07prb184441} we have this question numerically
investigated for the corresponding SU(3) model. We have found evidence
that the domain wall placed on adjacent sites as shown in
Fig.\,\ref{fig:3-exc}\,(b) is the elementary excitation.  As pointed
out above, this domain wall corresponds to an excitation transforming
according to the anti--fundamental representation $(0,1)$ of the SU(3)
model. This is what we expect as the spinon excitations for
antiferromagnetic SU(3) spin chains generally transform according to
the representation $(0,1)$. For the Haldane--Shastry model, this was
shown by explicit construction of the exact one spinon
eigenstates~\cite{schuricht-05epl987,schuricht-06prb235105}.  On more
general grounds, the low--energy behavior of an SU(3) spin chain with
fundamental representation is described by the SU(3)$_{k=1}$
Wess--Zumino--Novikov--Witten (WZNW) model. The elementary excitations
in this model transform likewise according to the representation
$(0,1)$ under SU(3)
rotations~\cite{bouwknegt-96npb345,schoutens97prl2608}.

\begin{figure}[h!]
  \centering
  \includegraphics[scale=0.6]{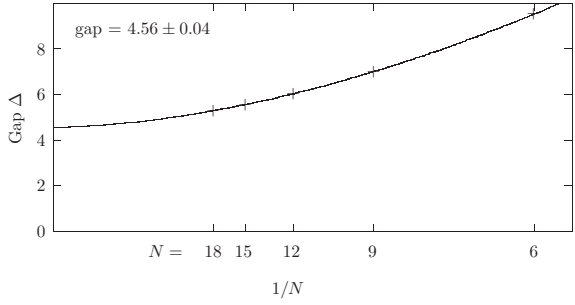}
  \caption{The gap size of a chain with certain length $N$ is plotted
    vs. the inverse chain length. Data points for $N=6$, $9$, $12$,
    $15$, and $18$ yield the fitted value of $\Delta=4.56 \pm 0.04$.}
  \label{fig:gap-scaling}
\end{figure}
Note that the triplet excitations are gaped, as it costs a finite
energy to break a trimer. In Fig.\,\ref{fig:gap-scaling} we have
plotted the gap size for the finite chains versus the inverse system
length. Fitting data points for $N=6$, $9$, $12$, $15$, and $18$
yields a huge gap $\Delta=4.56 \pm 0.04$ for the thermodynamic
limit. This gap, however, is not the Haldane gap exhibited by the
antiferromagnetic $S=1$ Heisenberg chain or the AKLT model, which we
understand as due to a confinement force between spinons.  This is
consistent, as the SU(3) Heisenberg model does not display a Haldane
gap~\cite{affleck-86lmp57,greiter-07prb184441}. 
The static spin--spin correlations of our model \eqref{ham.trimer}
decay abruptly, as adjacent trimer singlets are uncorrelated.

So far, we have used the mappings \eqref{SU3-SU2-mapping} and
\eqref{su3-permuation-operator} to construct a model of an
antiferromagnetic $S=1$ spin chain with an SU(3) symmetry. As a
possible generalization, we might look for models of spin $S$
antiferromagnets with an SU($2S+1$) symmetry, \ie models in which
$n=2S+1$ spins might form an $n$--mer rather than three neighboring
spins a trimer. Corboz \ea \cite{corboz-07prb220404} advocated that
such an $n$--merized phase should appear in an appropriate
$J_1$--$J_2$ model provided the ratio $J_1/J_2$ exceeds a certain
critical value. Note that this is not in contradiction to our previous statement that $n+1$--site interactions are required for tetramerization. Corboz \ea~proposed an $n$--merized phase which becomes exact in the thermodynamic limit only. Our statement is restricted to models describing exact $n$--merization.

The generalization of our construction is based on the mapping
\begin{equation}
  \label{sun-su2-mapping}
 \mathcal{P}^{(n)}_{\alpha\beta}~=~
 \sum_{\rho=0}^{n-1} a_{\rho}^{(n)} \left( \bs{S}_{\alpha}\bs{S}_{\beta} 
  \right)^{\rho},
\end{equation}
where the $\bs{S}_{\alpha}$ are spin $S$ operators and
$\mathcal{P}^{(n)}_{\alpha\beta}$ with $n=2S+1$ is the SU($n$)
symmetric permutation operator, exists in general.  A method to
determine the constants $a_{\rho}^{(n)}$ has been developed by
Kennedy~\cite{kenndy92jpa2809}; Itoi and Kato~\cite{itoi-97prb8295}
obtained them explicitly up to $S=2$. We now use these results to
introduce a spin $3/2$ parent Hamiltonian which describes
tetramerization.

\section{Model for tetramerization}
\begin{figure}[b]
  \centering
  \includegraphics[scale=0.67]{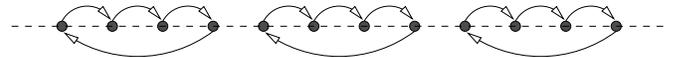}
  \caption{Illustration of a tetramer state on a chain with $N=4\mu$
    sites, where four neighboring spins are antisymmetrically
    coupled.}
  \label{fig:tetramer-state}
\end{figure}

We consider a spin $S=3/2$ antiferromagnetic chain with $N=4\mu$ sites
($\mu$ integer) and PBCs. The operator $Q[i,j,k,l]$ creates
a tetramer singlet on sites $i,j,k,$ and $l$,
\begin{equation}
  \label{tetramer-singlet}
\begin{split}
&Q[i,j,k,l]\vac=\\[5pt]
&\frac{1}{\sqrt{4!}}\sum_{\alpha,\beta,\gamma,\delta=\pi\left(\frac{3}{2},
\frac{1}{2},-\frac{1}{2},-\frac{3}{2}\right)}
c_{i,\alpha}^{\dagger}c_{i+1,\beta}^{\dagger}c_{i+2,\gamma}^{\dagger}
c_{i+3,\delta}^{\dagger}\vac.
\end{split}
\end{equation}
The sum in \eqref{tetramer-singlet} extends over all 24 permutations
$\pi$ of the four states
$\ket{\frac{3}{2}}_i=c_{i,\frac{3}{2}}^{\dagger}\ket{0}$,
$\ket{\frac{1}{2}}_i=c_{i,\frac{1}{2}}^{\dagger}\ket{0}$,
$\ket{-\frac{1}{2}}_i=c_{i,-\frac{1}{2}}^{\dagger}\ket{0}$, and
$\ket{-\frac{3}{2}}_i=c_{i,-\frac{3}{2}}^{\dagger}\ket{0}$. The four
tetramer ground states are given by
\begin{equation}\label{tetramer-groundstates}
 |\phi_{\nu}\rangle = \prod_{i=0}^{\mu-1} 
Q[4i+\nu,4i+1+\nu,4i+2+\nu,4i+3+\nu] \vac 
\end{equation}
with $\nu=1,2,3,$ or $4$. One of these ground
states is illustrated in Fig.~\ref{fig:tetramer-state}.

In analogy to the MG and the trimer model \eqref{ham.trimer}, a
Hamiltonian which annihilates the tetramer states can be written
as a sum over projection operators. For convenience we introduce
an auxiliary operator ${\tilde{\mathcal{X}}}^{(5)}$ acting on five
neighboring sites
\begin{equation}
\nn
    {\tilde{\mathcal{X}}}_i^{(5)}= \sum_{\underset{j<j'}{j,j' = i}}^{i+4}
    \mathcal{P}^{(4)}_{jj'}= \sum_{\underset{j<j'}{j,j' = i}}^{i+4}
    \sum_{\rho=0}^3 a_{\rho}^{(4)} \left(\bs{S}_{j}\bs{S}_{j'} 
  \right)^{\rho}
\end{equation}
with the constants~\cite{kenndy92jpa2809,itoi-97prb8295}
\begin{equation}
  \label{su4-constants}
  a_0^{(4)}=-\frac{67}{32}~,~a_1^{(4)}=-\frac{9}{8}~,
 ~a_2^{(4)}=\frac{11}{18}~,~a_3^{(4)}=\frac{2}{9}.
\end{equation}
Note that $\mathcal{P}^{(4)}_{jj'}$ is the SU(4) symmetric permutation
operator whereas the $\bs{S}_j$ are spin 3/2 operators.  In terms of
the auxiliary operator the parent Hamiltonian is given by
\begin{equation}
  \label{ham.tetramer}
  \mathcal{H}^{\textrm{quadr.}}=\sum_i \big( \,{\tilde{\mathcal{X}}}_i^{(5)}
   +5 \,\big)
  \big( \,  {\tilde{\mathcal{X}}}_i^{(5)}  + 2 \,\big).
\end{equation}
This Hamiltonian is positive semi--definite, exact by construction,
and annihilates the states \eqref{tetramer-groundstates}:
\begin{equation}\nn
  \big( \,{\tilde{\mathcal{X}}}_i^{(5)}+5 \,\big)\big( \,  
  {\tilde{\mathcal{X}}}_i^{(5)}  + 2 \,\big)
  |\phi_{\nu}\rangle=0,
\end{equation}
which implies $\mathcal{H}^{\textrm{quadr.}} |\phi_{\nu}\rangle=0$ for
$\nu=1,2,3,$ or $4$. It represents an exact model for
tetramerization in a spin 3/2 antiferromagnet. 
\begin{figure}[t!]
  \centering
  \includegraphics[scale=0.685,angle=0]{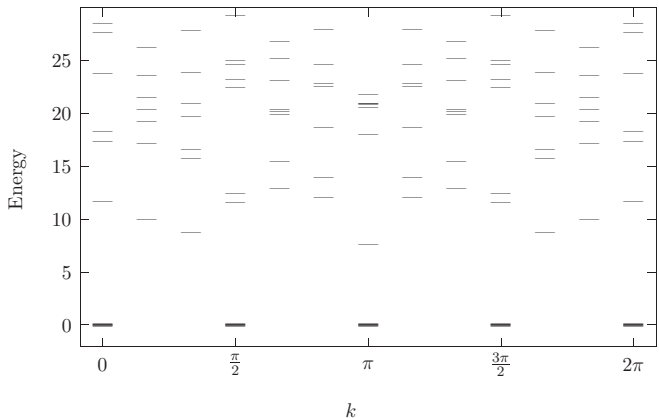}
  \caption{Spectrum of the spin $3/2$ model \eqref{ham.tetramer} for
    tetramerization on a chain with $N=12$ sites. The zero energy
    ground states (labeled by thick lines in the spectrum) are at
    $k=0,\frac{\pi}{2}, \pi, \frac{3\pi}{2}$ in the Brioullin zone.}
  \label{fig:spectrum32}
\end{figure}

We have confirmed our predictions with exact diagonalization (ED) for
chains with $N=8$ and $N=12$ sites. The spectrum of the 12 site chain
is shown in Fig.\,\ref{fig:spectrum32}, where the four zero--energy
ground states are located at $k=0$,$\frac{\pi}{2}$, $\pi$, and
$\frac{3\pi}{2}$, reflecting that the ground states are invariant
under translations by four lattice spacings.

Since the Hamiltonian \eqref{ham.tetramer} is similar to
\eqref{ham.trimer} , we contend ourselves with a brief discussion.
Interactions between five neighboring sites are required to ensure
annihilation of each tetramer singlet and, hence, annihilation of the
states \eqref{tetramer-groundstates}.  In order to find the correct
projection operators, we couple five SU(4) spins with fundamental
representation: $(1,0,0)\otimes(1,0,0)\otimes(1,0,0)\otimes
(1,0,0)\otimes(1,0,0)=4\cdot(1,0,0)\oplus5\cdot(0,1,1)\oplus
6\cdot(2,0,1)\oplus 5\cdot(1,2,0)\oplus 4\cdot(3,1,0)\oplus(5,0,0)$.
In the tetramer ground states, only the representations $(1,0,0)$ and
$(0,1,1)$ are present.  The projection operators onto the subspaces in
which the representations $(1,0,0)$ and $(0,1,1)$ are absent hence
annihilate the states \eqref{tetramer-groundstates}.  Since these
states are the only states which contain representations $(1,0,0)$ and
$(0,1,1)$ on five consecutive sites, a Hamiltonian in terms of
projection operators may be written as
\begin{equation}\nn
\begin{split}
  \mathcal{H}^{\textrm{quadr.}}&\!=
\!\!\!\sum_{\langle ijklm \rangle}\!\!\Bigg(
P \setlength{\unitlength}{5pt}
\begin{picture}(4,0)(0,0.5)
\linethickness{0.3pt}
\put(0,1){\line(1,0){3}}
\put(0,0){\line(1,0){3}}
\put(0,-1){\line(1,0){2}}
\put(0,-1){\line(0,1){2}}
\put(1,-1){\line(0,1){2}}
\put(2,-1){\line(0,1){2}}
\put(3,0){\line(0,1){1}}
\end{picture}(i,j,k,l,m)+
P \setlength{\unitlength}{5pt}
\begin{picture}(4,0)(0,0.5)
\linethickness{0.3pt}
\put(0,1){\line(1,0){3}}
\put(0,0){\line(1,0){3}}
\put(0,-1){\line(1,0){1}}
\put(0,-2){\line(1,0){1}}
\put(0,-2){\line(0,1){3}}
\put(1,-2){\line(0,1){3}}
\put(2,0){\line(0,1){1}}
\put(3,0){\line(0,1){1}}
\end{picture}(i,j,k,l,m)\Bigg.\\
\Bigg.+&P \setlength{\unitlength}{5pt}
\begin{picture}(6,0)(0,0.5)
\linethickness{0.3pt}
\put(0,1){\line(1,0){5}}
\put(0,0){\line(1,0){5}}
\put(0,0){\line(0,1){1}}
\put(1,0){\line(0,1){1}}
\put(2,0){\line(0,1){1}}
\put(3,0){\line(0,1){1}}
\put(4,0){\line(0,1){1}}
\put(5,0){\line(0,1){1}}
\end{picture}(i,j,k,l,m)+
P \setlength{\unitlength}{5pt}
\begin{picture}(5,0)(0,0.5)
\linethickness{0.3pt}
\put(0,1){\line(1,0){4}}
\put(0,0){\line(1,0){4}}
\put(0,-1){\line(1,0){1}}
\put(0,-1){\line(0,1){2}}
\put(1,-1){\line(0,1){2}}
\put(2,0){\line(0,1){1}}
\put(3,0){\line(0,1){1}}
\put(4,0){\line(0,1){1}}
\end{picture}(i,j,k,l,m)\Bigg),
\end{split}
\end{equation}
where $\langle \cdot \rangle$ indicates summation over five
neighboring sites along the chain. We then replace the projection
operators by SU(4) spin operators and rewrite the SU(4) spin operators
by spin $S=3/2$ operators using the mapping \eqref{sun-su2-mapping}.
This yields the Hamiltonian \eqref{ham.tetramer}.

Whereas the model for trimerization exhibited two possible candidates
for the low-lying excitation, the model for tetramerization exhibit
three candidates.  When breaking a tetramer singlet which we might
view as an SU(4) singlet, one obtains either an individual SU(4) spin
with fundamental representation $(1,0,0)$, or two antisymmetrically
coupled spins on adjacent sites with representation $(0,1,0)$, or
three antisymmetrically coupled spins on consecutive sites with
anti-fundamental representation $(0,0,1)$. Whereas the representations
$(1,0,0)$ and $(0,0,1)$ are four--dimensional, the representation
$(0,1,0)$ is six--dimensional.  The excitations with fundamental and
anti-fundamental representation correspond in case of the spin $S=3/2$
chain to the quadruplet excitation carrying spin $S=3/2$. However, if
one couples two spin $S=3/2$, the decomposition into irreducible
representations does not contain a six--dimensional representation:
$\bs{\frac{3}{2}}\otimes\bs{\frac{3}{2}}=
\bs{0}\oplus\bs{1}\oplus\bs{2}\oplus\bs{3}$. We may hence discard the
representation $(0,1,0)$ from the list of candidates.

\begin{figure}[h]
  \centering
\vspace{10pt}
  \includegraphics[scale=0.7]{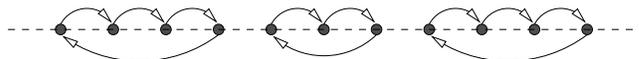}
  \caption{Quadruplet excitation serving as a domain wall between two
    of the tetramer ground states on a spin $S=\frac{3}{2}$ chain
    with $4\mu+3$ sites.}
  \label{fig:4bar-exc}
\end{figure}

In order to motivate that the quadruplet excitation placed on three
consecutive sites is the elementary excitation, we consider an SU(4)
spin chain. For the SU($n$) Haldane--Shastry model (HSM) (and
particularly for the SU(4) HSM), the lowest lying excitation carries
the anti--fundamental representation~\cite{schuricht-06prb235105}.
Moreover, the low--energy behavior of SU(4) spin chains with
fundamental representation is described by the SU(4)$_{k=1}$ WZNW model
where the lowest lying excitation carries the quantum number
$(0,0,1)$, \ie it transforms under the anti--fundamental
representation under SU(4)
rotations~\cite{bouwknegt-96npb345,schoutens97prl2608}. This
representation can be realized by antisymmetrizing three SU(4) spins
with fundamental representation $(1,0,0)$. It corresponds in the spin
S=3/2 model to the quadruplet excitation placed on three consecutive
sites (see Fig.\,\ref{fig:4bar-exc}).
This quadruplet excitation is gaped, as it costs a finite energy
to break a tetramer singlet. 

Most recently, the SU(4) $J_1$--$J_2$ model with fundamental
representation was investigated both within a level spectroscopy
analysis of the ED data~\cite{laeuchli:unpub} and within the density
matrix renormalization group (DMRG) analysis~\cite{rachel-:unpub}.
As proposed by Corboz \ea~\cite{corboz-07prb220404}, the existence
of a tetramerized phase was verified for the regime 
$J_1\approx J_2$~\cite{laeuchli:unpub,rachel-:unpub}.

\section{Conclusion}
In conclusion, we propose exact models for trimerization and
tetramerization in antiferromagnetic spin 1 and spin $3/2$ chains,
respectively.  They can be seen as generalizations of the
Majumdar--Ghosh model.  The models consist of a local Hamiltonian
involving four site or five site interactions, respectively, with a
three--fold or four--fold degenerate ground state.  The ground states
are products of local trimer or tetramer singlets where each trimer or
tetramer is placed on three or four consecutive sites. We have
numerically investigated the excitation spectrum and verified the
validity of both models using exact diagonalization on finite chains.

After this work was completed we became aware of another interesting
generalization~\cite{trebst-08prl050401} of the Majumdar--Ghosh model
where three--anyon interactions in the context of a chain of Fibonacci
anyons are investigated.

\section{Acknowledgements}
We would like to thank A. L\"auchli for a stimulating discussion.
SR acknowledges support from the Cusanuswerk.



\end{document}